**RESEARCH PAPER**     OPEN ACCESS

# MangngalApp- An integrated package of technology for COVID- 19 response and rural development: Acceptability and usability using TAM

**Billy S. Javier\*, Leo P. Paliuanan, James Karl A. Agpalza, Jesty S. Agoto**

*College of Information and Computing Sciences, Cagayan State University, Aparri, Philippines*

**Abstract**
The COVID19 pandemic has challenged universities and organizations to devise mechanisms to uplift the well-being and welfare of people and communities. In response, the design and development of an integrated package of technologies, MangngalApp- A web-based portal and mobile responsive application for rural development served as an opportunity. It showcases different packets of technologies that were outputs of R&D in the field of fisheries and aqua-culture, innovations that were IP-protected, and technologies that harness locally available resources for post-harvest development and aiding in sustaining growth and development in the communities. This paper focused on the usability and acceptability of the MangngalApp implementing a descriptive research design using the Technology Acceptance Model or TAM and ISO 25010 software quality standards. Constrained by government health restrictions due to COVID- 19, a Google form-based questionnaire was forwarded to consented participants via an email with the attached consent and evaluation form. Results revealed that the MangngalApp was found to be very acceptable and usable, and compliant to ISO 25010 software quality characteristics to the higher extent. From the results, it is concluded that the developed MangngalApp will be a usable and responsive technology that aids to rural development especially among target users- fishers, gatherers, processors, traders, and farmers. Considering compatibility and usefulness, the MangngalApp is expected to provide greater social development in the community.

**\*Corresponding Author:** Billy S. Javier ✉ billyjavier@csu.edu.ph

**Introduction**
The COVID 19 pandemic has disrupted many organizations, government and non-government institutions, schools, companies, and various communities. As a result, more displaced workers and job losses increased, more families sent home, locked down due to COVID19 restrictions and uncertain of how and where to obtain immediate income for the family. The government may have provided financial assistance to erring families and fed empty stomachs. However, resources deplete as no concrete measure to total stop the threat of the on-going pandemic the Filipino people is enjoying. The Cagayan State University is mandated to transforming the lives of people and communities through high quality instruction, innovative research, development, and production. Through the years, CSU has been working hard on innovating technologies that could help alleviate poverty, increase productivity and improve socioeconomic status of the communities, and help in sustaining and protecting the environment. However, no matter how promising these technologies are if these packages of technologies are not widely accessible to target communities, to its intended stakeholders: fishers, farmers, gatherers, and processors. In fact, Sharma A, and Kiranmayi, D (2019) was unable to find in many literature and studies pertaining to a package of technologies as an IEC initiative to adopting and utilizing research-based fisheries technologies, post-harvest technologies, and aquaculture techniques. Most of the 124 applications reported focused on mobile apps for angling, aquaculture management, aquarium management, marine fisheries, and fisheries governance, marketing and biology.

Research project generating innovative technologies and products has been funded and curated by experts in the various fields leading to technology commercialization. These then has to be extended to communities via available and relevant technologies so that as an academic institution, it really radiates its mantra of improving the lives of people and communities. The MangngalApp research program was generally geared at providing a solution for a well-informed utilization of the packets of technologies (POTs) developed as results of scientific inquiries and experiments of the University and collaborating agencies. It has been said that technologies should be utilized by the communities, adopted via technology-transfer, generating income from them. However, access to POTs may have not deliberately reaching the realms of coastal communities. Lack of or limited access to POTs among fishers, farmers, gatherers, and processors may cause inefficiency, increased cost for production, and lower productivity among fishers, fish processors and gatherers, as well as farmers in the coastal communities in northern Philippines.

With aqua-marine as banner program in the Aparri Campus, a multi-disciplinary research program was proposed with the hope of generating a package of technology showcasing the science-based packages of technologies of university along fishing activities, seaweed farming, post-harvest, product development and more. The research is expected to benefit the coastal communities through provision of mobile-ready and friendly application accessible to users aiding to improve productivity, increased awareness and protection for the environment, and providing livelihood for women and differently-able persons. Packages of technologies developed will be best adopted or utilized in the community once an integrated package or technology is made available. Hence, the potential benefits expand from the fishers in the conduct of and management of their fisheries activities to any other intended users. Coastal farmers will be able to uncover scientific ways to conservation and management of marine species or seaweeds. Fish processors will have the potential to improve productivity, creation of jobs, and increased revenues.

Adapting the vision of the Food and Agriculture Organization of the United Nations (FAO) on enhancing the role of small-scale fisheries in contributing to poverty alleviation and food security, the project also focused on understanding the technology awareness, technology adoption practices, the information needs and seeking behaviors, media literacy and media adoption of various stakeholders in the fishing communities of Northeastern Cagayan Philippines. In the academe, students and teachers may benefit from the having obtained the scientific packages of technologies for instruction purposes, and an opportunity for more relevant research formulation. The results of the study hope to provide and cultivate new knowledge for students, researchers, and teachers. In so doing, students and teachers may devise projects, programs, and studies that could add up to the packages of technologies. Institutions or organization may have devise appropriate strategies, programs and plans from data mining and knowledge data discovery thru the program.

The emergence of an information, communication, and education platform through varied technologies is a must especially in the dissemination of scientific results and innovations from rigid experiments and research. Digital visibility is considered an efficient and reasonable way to publicize the outputs of innovative developments and research results (Magdalinou, 2019). The Technology Acceptance Model (TAM) is a theory in information systems that explain how consumers come to embrace the use of a technology. When consumers are introduced with new technology, the model argues that a variety of factors influence their decision on how and when to use it. TAM has been critiqued for a variety of reasons, but it is a useful overall framework that is compatible with several studies examining the elements that influence older individuals' willingness to utilize new technology (Braun, 2013).

This paper generally aims to describe the usability and acceptability of the developed mobile responsive web project known as MangngalApp - an integrated package of technology using open-source web development platform.

The assessment of the usability and acceptability of the MangngalApp using the Technology Acceptance Model (TAM) focused on (a) Perceived Ease of Use, (b) Perceived usefulness, (c)Attitudes towards usage, (d) Behavioral intention to use, and Relevance to the present job. In addition, the assessment of the developed MangngalApp based on ISO 25010 software quality characteristics has been reported.

**Materials and methods**

The descriptive research design was implemented in this part of the project. The assessment of the usability and acceptability of the developed Mangngal App using the Technology Acceptance Model or TAM was participated by 200 non-technical respondents. These included fishers, farmers, fish processors, gatherers, and households involved in post-harvest. A listing of which was taken from municipal agriculture office thru communications. Meanwhile, the assessment of the 20 technical respondents applying ISO 25010 software quality standards, provided proof of the compliance in terms of compatibility, reliability, user-friendliness, security, portability, and functional suitability. The profile of the technical respondents is presented herein in table 1. The survey-questionnaire included some profile data of respondents, their assessment of the MangngalApp, and an optional remark or comment part. A consent form was part of the questionnaire, while prior presentation or orientation on its use was provided via Google Meet.

The researchers took the assistance of partner-students and community leaders handled by the team in the locality to share the MangngalApp project and guide intended users including those involved in actual fishing, post-harvest development, processing, gathering, as well as those who are trading. This is a COVID-19 initiative of the project team in order to gather sentiments and assessment of those greater users. On the other hand, the technical respondents were communicated formally requesting their expertise, and provided the team consent to participate in the assessment.

The respondents in the evaluation of the technical compliance, usability and acceptability standards using TAM included 10 industry practitioners, 10 ICT teachers with experiences in databases, web development and design, and programming.

**Table 1.** Profile of the Technical Respondents.

| Participants Classification | Male | Female | Total | % |
|---|---|---|---|---|
| Industry Practitioners | 6 | 4 | 10 | 50.0 |
| ICT Teachers | 5 | 5 | 10 | 50.0 |
| Area of Interest | | | | |
| Web Design | 2 | 3 | 5 | 25.0 |
| Web Programming | 3 | 2 | 5 | 25.0 |
| Databases | 2 | 2 | 4 | 20.0 |
| Programming | 2 | 2 | 4 | 20.0 |
| Networks | 2 | | 2 | 10.0 |
| Years of Relevant ICT Experience | | | | |
| 1 to 3 | 4 | 5 | 9 | 45.0 |
| 4 to 6 | 6 | 3 | 9 | 45.0 |
| More than 6 | 1 | 1 | 2 | 10.0 |

The participants were notified via email on their participation in the assessment. A brief orientation via Google Meet was conducted to provide them overview of the project. The project team provided the link of web based MangngalApp project. They were given at least 2 to 5 weeks to access the web project and were requested to fill out the evaluation forms via Google Forms. Treating the assessment of the Usability and Acceptability of the MangngalApp using the Technology Acceptance Model, the 4-point Likert scale was used: 1 being not acceptable and usable to 4 being very acceptable and usable. The MangngalApp web portal was developed applying the Design Science Research (DSR) for Information Systems. The Design Science Research creates and evaluates IT artifacts intended to solve identified organizational problems, (Peffers, 2007). Accessible thru http://cics-csuaparri.org.ph/mangngalapp, the XAMPP development framework was mainly used. XAMPP is a cross-platform development tool involving the use of the PHP scripting language, My SQL database engine, and Apache web service. Other tools used included CSS3, HTML5 and JavaScript.

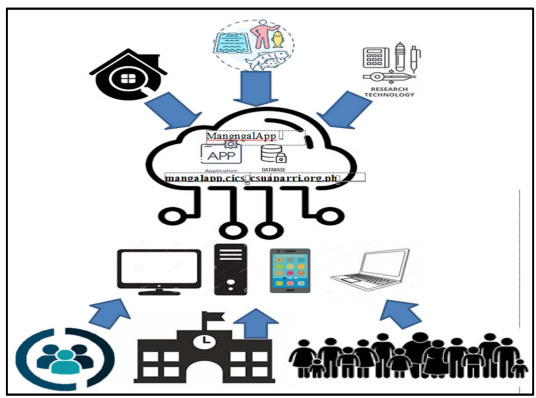

**Fig. 1.** The MangngalApp Ecosystem.

The MangngalApp Web Project is an ecosystem that involves people in the research and development, technologies for development and dissemination of research outputs, people and communities that are the main reason for this project towards rural development. The research outputs of the researchers and scientific organization that were IP-registered are highlighted for dissemination towards adoption strategy. Bridging the gap is maximizing the use of web tools and technologies that are accessible to the communities. The package of technology available in the current version contains 14 IP-registered technologies showcasing most of the CSU Aparri-based research and innovations.

Permissions were sought through the Knowledge and Technology Management Office and the Office of the Research and Development, and Extension. End-users of the project may click on the view process to see the detailed descriptions, as well as the steps involved in making, producing, or utilizing the technology. The project is scalable, it will still house other registered post-harvest technologies, fisheries-based products, and technologies supporting the different arrays of fisheries and aquaculture development for rural use.

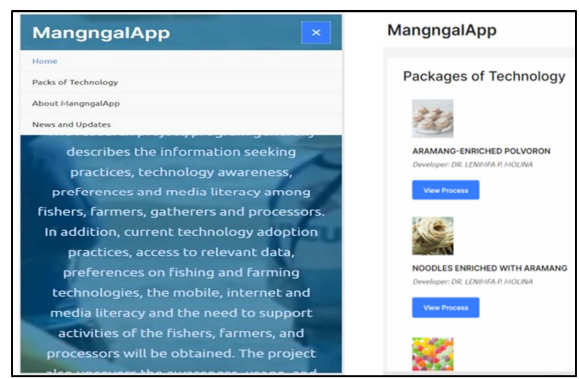

**Fig. 2.** Mobile View of the MangngalApp.

### Results and discussions

*Assessment of the Usability and Acceptability of the MangngalApp using the Technology Acceptance Model (TAM)*

Table 3 presents the results of the assessment made by the technical respondents along the aspects of perceived ease of use, perceived usefulness, attitudes towards usage, behavioral intention to use, and job relevance.

**Table 1.** Assessment of the Usability and Acceptability using TAM.

| Aspects of the Technology Acceptance | VAU f | AU f | SAU f | NAU f | Weighted Mean | DV |
|---|---|---|---|---|---|---|
| *Perceived Ease of Use* | | | | | *3.14* | *AU* |
| I feel that using MangngalApp would be easy for me | 8 | 12 | 0 | 0 | 3.40 | VAU |
| I feel that my interaction with MangngalApp would be clear and understandable | 6 | 12 | 1 | 0 | 3.25 | VAU |
| I feel that it would be easy to become skillful at using MangngalApp | 3 | 15 | 1 | 0 | 3.10 | AU |
| I would find MangngalApp to be flexible to interact with | 6 | 12 | 1 | 0 | 3.25 | VAU |
| Learning to operate MangngalApp would be easy for me | 6 | 12 | 2 | 0 | 3.20 | AU |
| It would be easy for me to get MangngalApp to do what I want to do | 2 | 13 | 4 | 0 | 2.90 | AU |
| I feel that my ability to determine MangngalApp ease of use is limited by my lack of experience | 4 | 11 | 3 | 2 | 2.85 | AU |
| *Perceived Usefulness* | | | | | *3.32* | *VAU* |
| Using MangngalApp in disseminating technologies to intended users would enable me or users to accomplish tasks more quickly | 11 | 8 | 1 | 0 | 3.50 | VAU |
| Using MangngalApp would improve my skills and is useful in the fishers and user's needs. | 7 | 12 | 1 | 0 | 3.30 | VAU |
| Using MangngalApp would increase my productivity | 7 | 12 | 1 | 0 | 3.30 | VAU |
| Using MangngalApp would enhance other users' capabilities adopting the technology shared. | 7 | 12 | 1 | 0 | 3.30 | VAU |
| Using MangngalApp would make it easier to know new technological updates in fishing, postharvest and related activities. | 6 | 12 | 2 | 0 | 3.20 | AU |
| I would find MangngalApp useful in helping the fishers and related sectors towards rural development. | 7 | 12 | 1 | 0 | 3.30 | VAU |
| *Attitudes towards Usage* | | | | | *3.43* | *VAU* |
| I believe it is a good idea to use the MangngalApp web project | 8 | 12 | 0 | 0 | 3.40 | VAU |
| I like the idea of using the MangngalApp web project | 8 | 12 | 0 | 0 | 3.40 | VAU |
| Using the MangngalApp is a positive idea | 10 | 10 | 0 | 0 | 3.50 | VAU |
| *Behavioural Intention to Use* | | | | | *3.22* | *AU* |
| I tend to use the MangngalApp web project for seeking new innovations in fisheries post-harvest and technologies. | 6 | 13 | 1 | 0 | 3.25 | VAU |
| I tend to use MangngalApp to enhance my interest in related fishing, aqua-culture, and post-harvest activities | 6 | 12 | 2 | 0 | 3.20 | AU |
| I tend to use the MangngalApp to provide multi-approaches on sharing and obtaining technological and innovations in fisheries, aqua-marine and post-harvest activities. | 6 | 12 | 2 | 0 | 3.20 | AU |
| Relevance of the MangngalApp to Current Job | | | | | *3.35* | *VAU* |
| In disseminating new packets of technologies along fisheries and aqua-marine, the usage of MangngalApp is important | 8 | 11 | 1 | 0 | 3.35 | VAU |
| In disseminating new packets of technologies along fisheries and aqua-marine, the usage of MangngalApp is timely relevant | 8 | 11 | 1 | 0 | 3.35 | VAU |
| Overall Weighted Mean | | | | | 3.29 | VAU |
| 3.25 – 4.00 >> Very acceptable and usable (VAU) | | 1.75 – 2.49 >> Somewhat acceptable and usable (SAU) | | | | |
| 2.50 – 3.24 >> Acceptable and usable (AU) | | 1.00 – 1.74 >> Not acceptable and usable (NAU) | | | | |

With an overall mean of 3.29, the assessment of the MangngalApp along the usability and acceptability aspects were found to be "very acceptable and usable" (table 3). Specifically, the assessment of perceived usefulness (3.32), their attitude towards usage (3.43), and relevance (3.45) were rated very acceptable and usable. The perceived usefulness could be associated to their perceived attitude towards its usage as well as how relevant the MangngalApp web project specially to intended users. For the purpose of clarity and understanding, the project team intended to have the MangngalApp project be assessed by the fishers, processors, farmers, traders, and gathers. However, the team was constrained to do the actual demonstration due to restrictions of the COVID-19 virus and high-risk alert levels of cases in the locality. The team also tried to meet the all intended participants via virtual setup in a video conferencing tool as well as used other strategies like communicating with students and leaders in the area. Feed backs from the students who were parents of the fishers and farmers as well as processors; said most of their parents prefer to have the project demonstrated in face-to-face setup so they could easily grasp the technology. The team decided to conduct the actual dissemination and training in the actual users in the ground upon notice of approval from relevant office still confirming to minimum health protocols. It is one of the key future directions the team is looking forward.

As presented, the group of non-technical respondents generally assessed the usability and acceptability of the MangngalApp as "very acceptable and usable" with a mean of 3.39. This rating is associated to the very acceptable and usable descriptive values for perceived usefulness, attitude towards usage, and job relevance. Interestingly, more male respondents perceived higher valuation of the Mangngal App compared to their female counterparts. Meanwhile, the technical respondents rated the aspects of TAM as "acceptable and usable" with a mean of 3.21. Higher assessment has been made by female industry practitioners with a mean of 3.38, especially along usefulness, attitudes towards usage, behavioral intention to use and job relevance. There were 40 percent of the respondents who rated the MangngalApp as overall very acceptable and usable.

**Table 2.** Detailed presentation of the assessment of the usability and acceptability

| Aspects of TAM | Technical Respondents | | | | Non-Technical Respondents | | | |
|---|---|---|---|---|---|---|---|---|
| | Male | Female | Weighted Mean | Descriptive Value | Male | Female | Weighted Mean | Descriptive Value |
| 1. Perceived ease of use | 3.07 | 3.14 | 3.10 | AU | 3.23 | 3.11 | 3.17 | AU |
| 2. Perceived usefulness | 3.00 | 3.50 | 3.27 | VAU | 3.47 | 3.27 | 3.37 | VAU |
| 3. Attitude towards usage | 3.33 | 3.50 | 3.40 | VAU | 3.47 | 3.47 | 3.47 | VAU |
| 4. Behavioral intention to use | 3.00 | 3.50 | 3.20 | AU | 3.40 | 3.07 | 3.23 | AU |
| 5. Job Relevance | 3.17 | 3.50 | 3.30 | VAU | 3.40 | 3.40 | 3.40 | VAU |
| Overall | 3.12 | 3.38 | 3.32 | AU | 3.37 | 3.23 | 3.30 | VAU |
| | | 3.21 | | AU | | | 3.39 | VAU |
| Percentage of those who rated the MangngalApp as overall "very acceptable and usable" | | | 40% | | | | 40% | |

Compliance to ISO 25010 software quality characteristics of the developed MangngalApp

**Table 3.** Summary table of the assessment of the developed MangngalApp based on ISO 25010 software quality characteristics.

| Indicator | Technical Evaluators | | Non-Technical (Fisher) | | Overall | |
|---|---|---|---|---|---|---|
| | WM | DV | WM | DV | WM | DV |
| Accuracy | 3.47 | VHE | 3.87 | VHE | 3.67 | VHE |
| Reliability | 3.53 | VHE | 3.90 | VHE | 3.72 | VHE |
| Security | 3.50 | VHE | 4.00 | VHE | 3.75 | VHE |
| Functional Suitability | 3.60 | VHE | 3.93 | VHE | 3.77 | VHE |
| Portability | 3.67 | VHE | 3.87 | VHE | 3.77 | VHE |
| Usability | 3.60 | VHE | 3.9 | VHE | 3.75 | VHE |
| Maintainability | 3.57 | VHE | 3.87 | VHE | 3.72 | VHE |
| Efficiency | 3.60 | VHE | 3.90 | VHE | 3.75 | VHE |
| Overall Weighted Mean | 3.57 | VHE | 3.91 | VHE | 3.74 | VHE |

Legend:

WM– Weighted Mean; DV– Descriptive Value

3.25-4.00 >> Very High Extent (VHE, 1.75-2.49 >> Fair Extent (FE)

2.50-3.24 >> High Extent (HE), 1.00-1.74>> Poor Extent (PE)

Presented in table the summary table of the assessment of the MangngalApp web project following the ISO 25010 software quality characteristics. The assessment of the technical and non-technical respondents revealed an overall remark of excellent with an overall mean of 3.74. Notably, both groups made a high remark or excellent highlighting functionality and portability aspects.

The functionality can be associated to the fact that the MangngalApp follows a WYSWYG approach making ease of access and functional. Meanwhile, the portability aspect could be associated to the project being compatible to varied devices making it convenient to users.

The participants were asked about their problems and challenges associated to the use of the MangngalApp. Although the participants are technical evaluators, it is believed that common issues will be experienced by the intended users. This includes but not limited to:

a. Internet connectivity issues
b. Not very good using via tablets PC
c. Limited contents only focused to fisheries and aquaculture
d. Cannot visualize from just an image

There were comments and suggestions highlighted by the respondents. This includes but not limited to:

a. Strengthen internet connection in the area
b. Share more techno guides that are easily understood by intended users
c. Produce video of the steps which are visibly understood by intended users
d. Add more contents not only along post-harvest and processing.
e. Translation of contents to Filipino or vernaculars if possible

Moreover, the overall impressions made by the participants include:

a. MangngalApp as a good project for rural development
b. The project is impressive
c. Great project especially if with more contents for the intended users
d. Very good one-stop IEC mechanism

Considering the above-mentioned, the project team is looking way forward to scale up the project, fast-track the translation to Filipino, as well as integrating other technologies that would benefit the communities for rural development. The translation is in coordination with owners of the technology.

**Conclusions**

The MangngalApp project was found to be very acceptable and usable based on the assessment of the technical respondents. There were uncontrolled issues or problems in the use of the MangngalApp, the constructive comments and suggestions, as well as the overall impressions over the project. Based on the ISO 25010 software quality characteristics, the respondents generally remark it as "excellent" with an overall mean of 3.74.

From the results, it is concluded that the developed MangngalApp will be a usable and responsive technology that aids to rural development especially among target users- fishers, gatherers, processors, traders, and farmers. Considering compatibility and usefulness, the MangngalApp is expected to provide greater social development in the community.

*Social Implications*

The use of the MangngalApp would offer greater opportunity for local users to livelihood development adopting the technologies being shared from the output of scientific undertakings at the University and with collaborators. Meanwhile, the adoption of the technologies may be undertaken providing opportunities for small to medium organizations towards livelihood development – forging partnership with the University and other stakeholders and private institutions.

*Project Limitations*

The researchers acknowledge the technical challenge that may have encountered by the participants as there were very limited face-to-face presentations made with intended users, thus may affect the results in the study. There is a need to perform actual demonstration with them upon approval of authorities and observing minimum health protocols.

*Recommendations*

From the results of the study, it is recommended to integrate the fully translated content and additional technologies geared towards full utilization of the MangngalApp especially creating opportunities for livelihood development. Further, the conduct of extension activities to adopt and utilize the project accessible in the web is highly encouraged thru demonstration activities forging collaboration with fishers and women organizations. In addition, there is a need to constantly update and make the project scalable providing other opportunities for rural development in general especially when new innovations are IP-registered from the research innovations in fisheries and aqua-marine. The development of a video production is suggested for actual demonstration of the processes involved especially in post-harvest or product development.

**Acknowledgement**
The research project would not be a success without the support of the administration of the Cagayan State University headed by Dr. Urdujah G. Alvarado, the kind assistance and support of the RDE for the funding thru VP for RDE Dr. Junel Guzman, as well as the commitment and leadership of the Campus Executive Officer Dr. Simeon R. Rabanal, Jr. The project team is ever grateful for the usual and unparalleled support and drive of the Coordinator for Research and Development Dr. Lenimfa Molina for sharing the technologies and helping us in the project contents. Special mention to Ms. Eunice Daluddung for her patience and assistance to the project team. Kind appreciation is extended to Dr. Corazon T. Talamayan for supporting us in the project. Morever, the assessment of the project as well as how could we better improve the MangngalApp is greatly attributed to the self-less sharing of time, effort and expertise of the industry practitioners and ICT teachers despite being very busy also. To all the fishers, farmers, processors, gatherers, and small-scale merchants – we owe this project to you, as our inspiration of doing the project towards rural development. Special mention goes to the member of the review committee in the 2 in-house reviews conducted – Engr. Gil Mark Hizon of DOST RO2 and Dr. Emma Ballad of BFAR RO2 for their constructive comments, guidance and inspiration: GAD-Focal Person Prof Kristine Lara, Extension coordinator Josie Bas-ong and KTM Coordinator Dr. Gilbert Magulod Jr for the inputs and support.